\documentclass[manuscript]{aastex}

\usepackage{longtable,lscape}
\usepackage{graphicx}

\shorttitle{The young open cluster Berkeley~55}
\shortauthors{Negueruela \& Marco}

\begin{document}


\title{The young open cluster Berkeley~55\altaffilmark{*}}
\altaffiltext{*}{Partially based on observations collected at the Nordic Optical Telescope and the William Herschel Telescope (La Palma)}


\author{Ignacio Negueruela\altaffilmark{1}, Amparo Marco\altaffilmark{1}\vspace{5mm}}

\affil{$^{1}$Departamento de F\'{i}sica, Ingenier\'{i}a de Sistemas y
Teor\'{i}a de la Se\~{n}al, Universidad de Alicante, Apdo. 99, E-03080
Alicante, Spain\email{ignacio.negueruela@ua.es}\email{amparo.marco@ua.es}}


\begin{abstract}
We present $UBV$ photometry of the highly reddened and poorly studied open cluster Berkeley~55, revealing an important population of B-type stars and several evolved stars of high luminosity. Intermediate resolution far-red spectra of several candidate members confirm the presence of one
F-type supergiant and six late supergiants or bright giants. The
brightest blue stars
are mid-B giants. Spectroscopic and photometric analyses indicate an age $50\pm10$~Myr. The cluster is located at a distance $d\approx4$~kpc, consistent with other tracers of the Perseus Arm in this direction. Berkeley~55 is thus a moderately young open cluster with a sizable population of candidate red (super)giant members, which can provide valuable information about the evolution of intermediate-mass stars.
\end{abstract}




\section{Introduction} \label{introduction}

After exhaustion of H in their cores, stars evolve toward lower $T_{{\rm eff}}$, becoming, according to their masses, red giants or supergiants (RSGs). Both high- and intermediate-mass stars are subject to complex physical processes in their later evolution, which result in important changes in their observable characteristics ($T_{{\rm eff}}$, $L_{{\rm bol}}$). As a consequence, their evolutionary tracks trace loops in the HR diagrams \citep[e.g.,][]{chiosi}. The shape of these loops depends on the physics of the stellar interior, generally modeled via poorly understood parameters \citep[e.g.,][]{chiosi,mf94,salas99,mm00}. The extent of semi-convection and overshooting has very important consequences on many aspects of stellar evolution, most notably the ratio of initial mass to white dwarf mass \citep[e.g.,][]{jeff97,weide00} and the boundary between stars that leave white dwarfs as remnants and those that explode as supernovae \citep[SNe; e.g.,][]{poel08}.  

M-type supergiants represent the final evolutionary stage of moderately massive stars, with typical initial masses in the 8--25$\,M_{\odot}$ range \citep{levesque05}.
Such objects are the immediate progenitors of Type II-P SNe, the most frequent type of supernova explosion in the Local Universe \citep{smartt09,smith11}. Most of the explosions come from low-mass RSGs, stars with initial masses $M_{*}\la12\,M_{\sun}$. The lower mass limit for a star to produce a supernova has been estimated at $\approx 8.5^{+1}_{-1.5}\,M_{\sun}$ \citep{smarttal}. Nevertheless, stars with lower masses are also classified as supergiants, generally K\,Ib objects, showing that the morphological separation between red supergiants and red giants does not coincide exactly with the boundary between high- and intermediate-mass stars.   

Since stars become more luminous as they experience blue loops and later enter the AGB branch, a given star may appear first as a bright giant and later in its evolution as a supergiant. Open clusters with large populations of evolved stars can help constrain the inputs of models and therefore improve our understanding of such basic questions. Identification and study of such clusters is thus an important astrophysical issue.

To this aim, we have analyzed a sample of poorly-studied,
cataloged, optically-visible clusters in the Northern Hemisphere, using 2MASS data \citep{skru06}. We made use of the $Q_{{\rm IR}}$ index, a reddening-free parameter, defined as $(J-H)-1.70(H-K_{{\rm S}})$, which has been proved to be a very useful tool for identification of intrinsically blue stars \citep[see, e.g.,][]{cp05,ns07} and separation of luminous red stars from red dwarfs \citep{neg11}. Berkeley~55 (Be~55) was readily identified as containing an obvious clump of red luminous stars associated with a sequence of intrinsically blue
stars. 

Be~55 is a faint, compact open cluster in the
constellation Cygnus. The WEBDA database\footnote{At {\tt http://www.univie.ac.at/webda/}} provides coordinates RA:~21h 16m 58s, Dec:~$+51^{\circ}\:45\arcmin\:32\arcsec$ ($\ell=93\fdg03$, $b=+1\fdg80$).
\citet{macie07} presented $BV$ photometry of an extended field around
the obvious cluster core and found it to be
extremely compact, with $r_{{\rm core}}=0\farcm7$. They derived a distance $d=1.2$~kpc and an age $\log t=8.5$ (315~Myr), which was later used by \citet{tadross} to calibrate a fit to the 2MASS data in the region. However, the cluster looks too compact for this age and distance, while the isochrone fit of \citet{tadross} would suggest that most of the evolved cluster members are AGB stars, an extremely unusual situation. 

In this paper, we present $UBV$ photometry of the cluster area and a spectroscopic survey of likely members, showing that its population is much younger than implied by the single-color fit of \citet{macie07}. From our new data, we show that Be~55 is a young ($\log
t\approx7.7$) open cluster with a rich population of evolved stars.

\section{Observations and data} 
\label{observations}

\subsection{Optical photometry}

 $UBV$ photometry of Be~55 was obtained in service mode using ALFOSC on the Nordic Optical Telescope at the Roque de los Muchachos Observatory (La Palma, Spain) on the night of June 14th, 2010. In imaging mode, the camera covers a field of $6\farcm5 \times 6\farcm5$ and has a pixel scale of $0\farcs19$/pixel.

One standard field from the list of \citet{landolt92}, PG~1323$-$086, was observed twice during the night in order to provide standard stars for the transformation. Since the number of measurements was too low to trace the extinction during the night, we used the median extinction coefficients for the observatory, after verifying that the two measurements of the standards were fully compatible with these values.
The images were processed for bias and flat-fielding corrections with the standard procedures using the CCDPROC package in IRAF\footnote{IRAF is distributed by the National Optical Astronomy Observatories, which are operated by the Association of Universities for Research in Astronomy, Inc., under cooperative agreement with the National Science Foundation}. Aperture photometry using the PHOT package inside DAOPHOT (IRAF, DAOPHOT) was developed on these fields with the same aperture, 21 pixels, for each filter.

Images of Be~55 were taken in 2 series of different exposure times to obtain accurate photometry for a magnitude range. The log of observations is presented in Table~\ref{photlog}. The reduction of the images of Be~55 was done with IRAF routines
for the bias and flat-field corrections. Photometry was obtained by
point-spread function (PSF) fitting using the DAOPHOT package
\citep{stetson1987} provided by IRAF. The apertures used are of the order of the FWHM. In this case, we used a value of 5 pixels for all images in the $U$ and $B$ filters and a value of 4 pixels for the $V$ filter images.

\begin{table}
\caption{Log of the photometric observations taken at the NOT on June 2010 for Berkeley 55.\label{photlog}}
\centering
\begin{tabular}{c c c}
\hline\hline
Berkeley 55 & RA = 21h 16m 59.9s & DEC = $+51\,^{\circ}$ $45\arcmin$ $59\farcs0$ \\
&(J2000)&(J2000)\\
\end{tabular}
\begin{tabular}{c c c}
\hline\hline
&\multicolumn{2}{c}{Exposure times (s)}\\
Filter & Long times & Short times \\
\hline
$U$ & 900 & 250\\
$B$ & 200 & 60 \\
$V$ & 40& 10 \\
\hline
\end{tabular}
\end{table}

In order to construct the PSF empirically,
we automatically selected bright stars (typically 25 stars). After this,
we reviewed the candidates and discarded those that did not fulfill all the
best conditions for a good PSF star. Once we had the list of PSF
stars ($\approx 20$), we determined an initial PSF by fitting the best
function between the 5 options offered by the PSF routine inside
DAOPHOT. We allowed the PSF to be  variable (in order 2) across the
frame to take into account the systematic pattern of PSF variability
with position on the chip.  
 
We needed to perform aperture correction for each frame in all filters.
Finally, we obtained the instrumental magnitudes for all stars. Using
the standard stars and the median extinction coefficients for the observatory, we carried out the transformation of the instrumental magnitudes to the standard system by means of the PHOTCAL package inside IRAF. 

\begin{figure*}[ht]
\resizebox{10 cm}{!}{\includegraphics{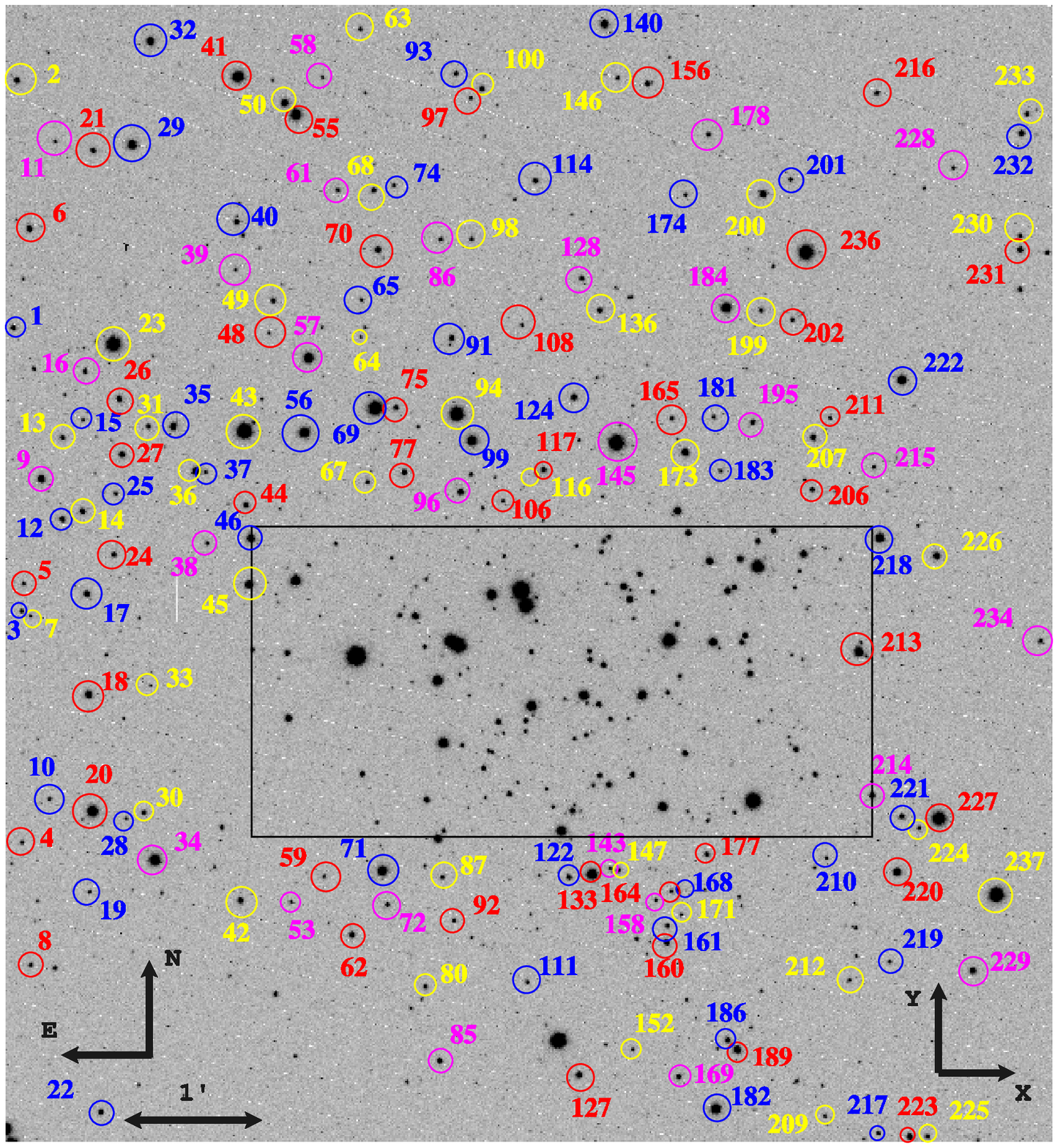}}
\centering
\caption{Finding chart for stars with photometry in the field of Berkeley 55. The image is one of our $V$-band frames. $XY$
  positions are listed in Table~\ref{allpos}, where they are
  correlated to RA and DEC. Stars inside the rectangle (which approximately defines the cluster core) are marked in Fig.~\ref{findercentre}. Each star is identified by the nearest marker in the same color as the circle around it. \label{findermain} } 
\end{figure*}

\begin{figure*}[ht]
\resizebox{16 cm}{!}{\includegraphics{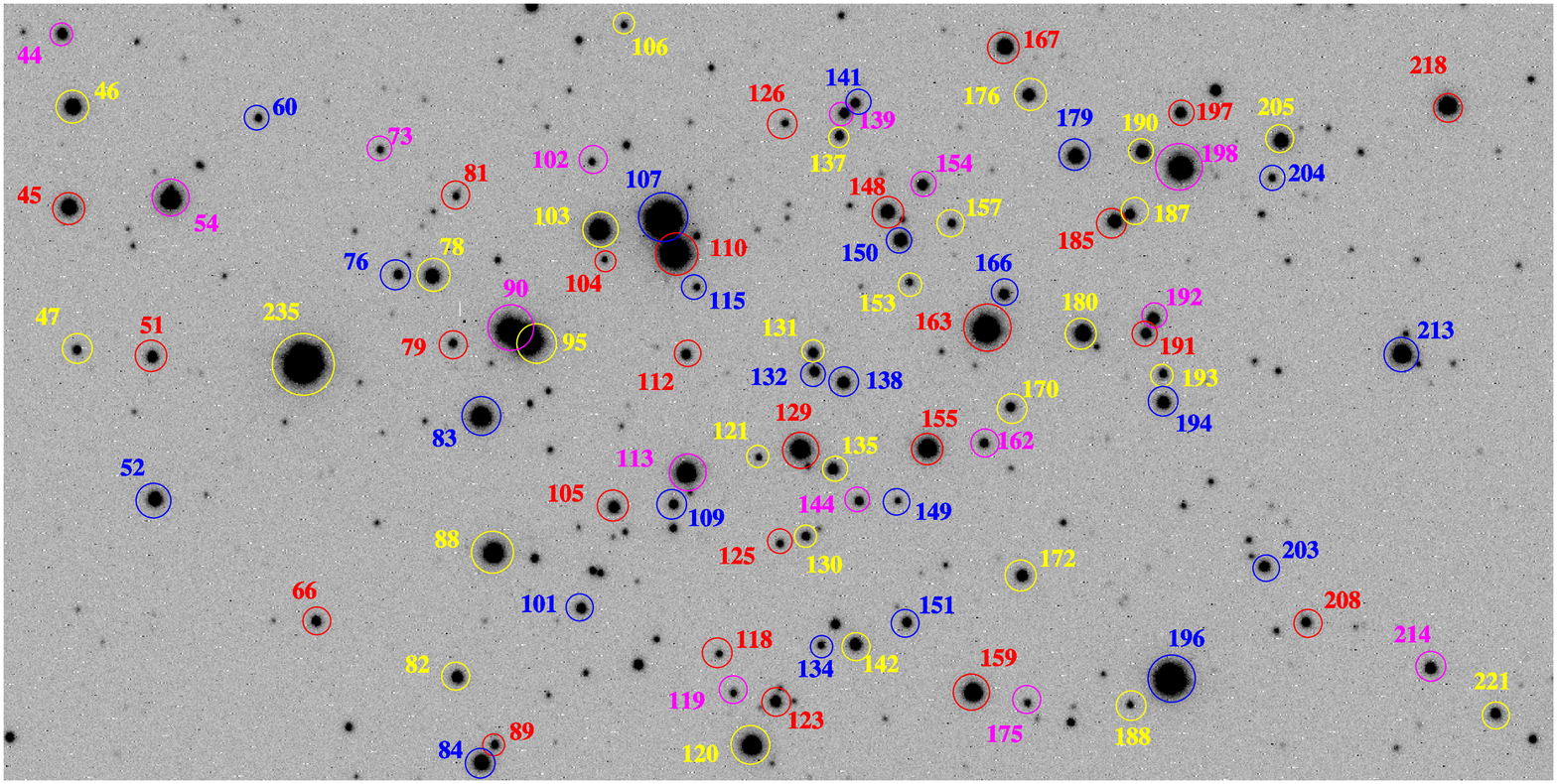}}
\centering
\caption{Finding chart for stars with photometry in the central part of Berkeley 55. The image is one of our $V$-band frames. $XY$
  positions are listed in Table~\ref{allpos}, where they are
  correlated to RA and DEC. Each star is identified by the nearest marker in the same color as the circle around it.\label{findercentre}} 
\end{figure*}

The number of stars that we could detect in all filters is limited by the long exposure time in the $U$ filter. We identify all stars with good photometry in all three filters on the image in Figures~\ref{findermain} and~\ref{findercentre}. In Table~\ref{allpos}, we list their $X$ and $Y$ positions as seen in Fig.~\ref{findermain}, and their identification with objects in the 2MASS catalog. The designation of each star is given by the number indicated on the images (Fig.~\ref{findermain} and Fig.~\ref{findercentre}). We have photometry for 237 stars in the field. In Table~\ref{allphot}, we list the values of $V$, $(B-V)$ and $(U-B)$ with the standard deviation and the number of measurements for each magnitude or index.

We can compare our photometry with the only existing dataset, that of \citet{macie07}. We find average differences (literature $-$ this work) in $V$ of $-0.1\pm0.3$. This shows important dispersion, but no systematic effects. The average difference in $(B-V)$ is $0.16\pm0.08$, suggesting a small systematic difference, though not much higher than the standard deviation. We note that, with a pixel size of $2\farcs2$ and a reported typical seeing around $5\arcsec$, the photometry of \citet{macie07} may be subject to important crowding effects near the cluster core. 

\subsection{2MASS data}

We obtained $JHK_{{\rm S}}$ photometry from the 2MASS catalog. The completeness limit of this catalog is set at $K_{{\rm S}}=14.2$. None of the stars analyzed is bright enough to come close to the non-linear regime for 2MASS. We used the 2MASS data to select targets for spectroscopy. 

\begin{figure}
   \resizebox{\columnwidth}{!}
   {\includegraphics[clip]{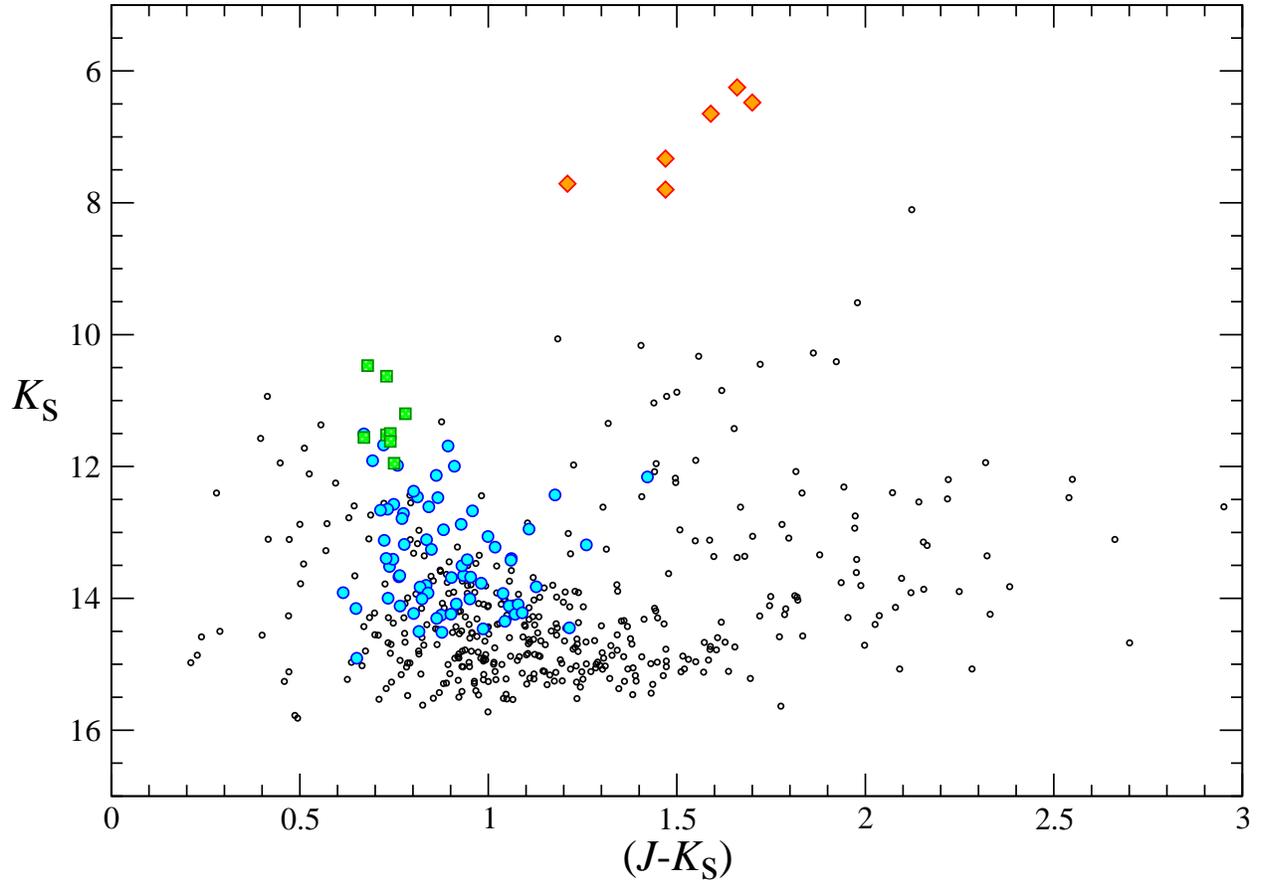}}
   \caption{$K_{{\rm S}}/(J-K_{{\rm S}})$ diagram for all stars in 2MASS within $3\arcmin$ of the arbitrarily defined center of Be~55. Stars selected as early-type stars based on their $Q_{{\rm IR}}$ index are marked as filled blue circles or green squares (if observed spectroscopically). The orange diamonds identify the clump of F--K supergiants. \label{fig:raw}}
   
    \end{figure}

We selected our targets using 2MASS $JHK_{{\rm S}}$ data. We took a   
circle of radius $3\arcmin$ around the cluster center and built the corresponding $K_{{\rm S}}/(J-K_{{\rm S}})$ diagram (displayed in Fig.~\ref{fig:raw}). All the stars with $K_{{\rm S}}\leq8.0$ (shown as diamonds) are clumped together in a small region of the diagram. Their $Q_{{\rm IR}}$ is $\approx0.3$ in all cases. This shows that they are not
field red dwarfs or red clump giants, and suggests that they may be luminous red stars \citep{neg11}. We also used the $Q_{{\rm IR}}$ index to separate early-type stars \citep[cf.][]{cp05,ns07}. As they are expected to show $Q_{{\rm IR}}\approx0.0$, we select stars with $-0.15\leq Q_{{\rm IR}}\leq0.08$ (shown as filled circles and squares in Fig.~\ref{fig:raw}). This range is intended to account for the typical errors in 2MASS (generally $0.03-0.05$~mag in a given color for stars with $K_{{\rm S}}=12-13$) and also include emission-line stars, which typically have $Q_{{\rm IR}}\la-0.05$ \citep[e.g.,][]{neg07}. This selection clearly shows all early-type stars in the field to concentrate around $(J-K_{{\rm S}})\sim0.8$, forming an approximately vertical sequence with a sharp blue edge at $(J-K_{{\rm S}})\sim0.7$ (Fig.~\ref{fig:raw}).

We performed the same analysis for 2MASS data on concentric circles of
increasing radii around the same position, finding no significant
increase in the number of candidate members. This is in good agreement
with the estimation by \citet{macie07} of a very concentrated cluster
with  $r_{{\rm core}}=0\farcm7$. Extending the search to $6\arcmin$,
we only find one candidate that seems 
to fall together with the red clump, both in the $Q/K_{{\rm S}}$ and
$(J-K_{{\rm S}})/K_{{\rm S}}$ diagrams. This is 2MASS
J21165198+5150410, which we consider a candidate member below (under the name S61). 

\subsection{Spectroscopy}

We obtained spectra in the far red of the brightest early-type candidates (identified as green squares in Fig.~\ref{fig:raw}) and the seven candidate red luminous stars using the red arm of the ISIS double-beam 
spectrograph, mounted on the 4.2-m William Herschel Telescope (WHT) in La
Palma (Spain). The instrument was fitted with the R600R grating and 
the {\it Red+} CCD. This configuration covers the
7600\,--\,9000\AA\ range in the unvignetted section of the CCD with a nominal dispersion of 0.5\AA/pixel.
Observations were taken during a service night on 26th July 2007 and
then completed during a run on 21st August 2007. In July, the CCD was unbinned, and a $1\farcs5$ slit was used. In August, the CCD was binned by a factor 2 in the spectral direction, and a  $1\farcs2$ slit was used. In both cases, the resolution element is expected to be $\sim$4~unbinned pixels. This has been checked by measuring the width of arc lines, which is on average $\approx2.1$\AA\ for both configurations. The resolving power of our spectra is therefore $R\sim4\,000$.

A log of these observations is presented in Table~\ref{logspec}, where each star is given an identification starting with S, and the correspondence with the numbering in the photometry is indicated. 
In addition to the stars targeted, two other photometric members also fell inside the slit by chance. One of them, \#213 (S19 in Fig.~\ref{fig:bestars}), is a bright member which turns out to be a Be star. The other one, \#170, is very faint and the spectrum is noisy, simply confirming that it is a B-type star.

\begin{table}
  \caption{Log of spectroscopic observations.\label{logspec}}
  \begin{tabular}{lccccc}
  \tableline
  \tableline
   Star   & Phot\tablenotemark{1} & Exposure &Date & Time & SNR\tablenotemark{2}    \\
& & Time (s) & & (UT) & \\
 \tableline
S1\tablenotemark{3}  &  $-$& 200 & 21 Aug 2007 & 01:07&  300\\
S2  & 196 & 400 & 26 Jul 2007 & 00:43&  340\\
S3  & 163 & 400 & 26 Jul 2007 & 00:43&  220\\ 
S4  & 110 & 400 & 26 Jul 2007 & 01:12&  260\\       
S5  & 107 & 400 & 26 Jul 2007 & 01:12&  310\\        
S6  & 145 & 400 & 26 Jul 2007 & 00:58&  330\\       
S61 & $-$ & 200 & 21 Aug 2007 & 01:13&  370\\
\tableline
S7  & 94 &  400 & 26 Jul 2007 & 01:12&  140\\ 
S8  & 99 & 1000 & 21 Aug 2007 & 01:27&  110\\
S9  & 198 & 400 & 26 Jul 2007 & 00:58&  110\\ 
S10 & 83 &1000 &  21 Aug 2007 & 01:27& 140\\
S12 & 113 &600 &  26 Jul 2007 & 01:24& 90\\
S13 & 129 &600 &  26 Jul 2007 & 01:24& 90\\
S15 & 159 &1000 &  21 Aug 2007 & 02:01& 90\\
S16 & 133 &1000 &  21 Aug 2007 & 02:01& 100\\
\tableline
\end{tabular}
\tablenotetext{1}{Cross-correlation with numbering system in Table~\ref{allphot}.}
\tablenotetext{2}{ The SNR has been estimated for the B-type stars from the RMS of fits to continuum regions. In the red supergiants, where there are no spectral windows free of lines, it is an extrapolation based on the count rate. The SNR is for each 0.5\AA\ binned pixel.}
\tablenotetext{3}{2MASS~J21165840+5143262, saturated in our photometry.}
\end{table}


All spectroscopic data were reduced using the Starlink software
packages CCDPACK \citep{Draper2000} and FIGARO \citep{shortridge}. We used standard procedures for bias subtraction and flat-fielding (with internal halogen lamps). The spectra have been normalized to the continuum using DIPSO \citep{Howarth1998}.

\section{Results}

\subsection{Observational HR diagram}

\begin{figure}
   \centering
   \includegraphics[width=\columnwidth]{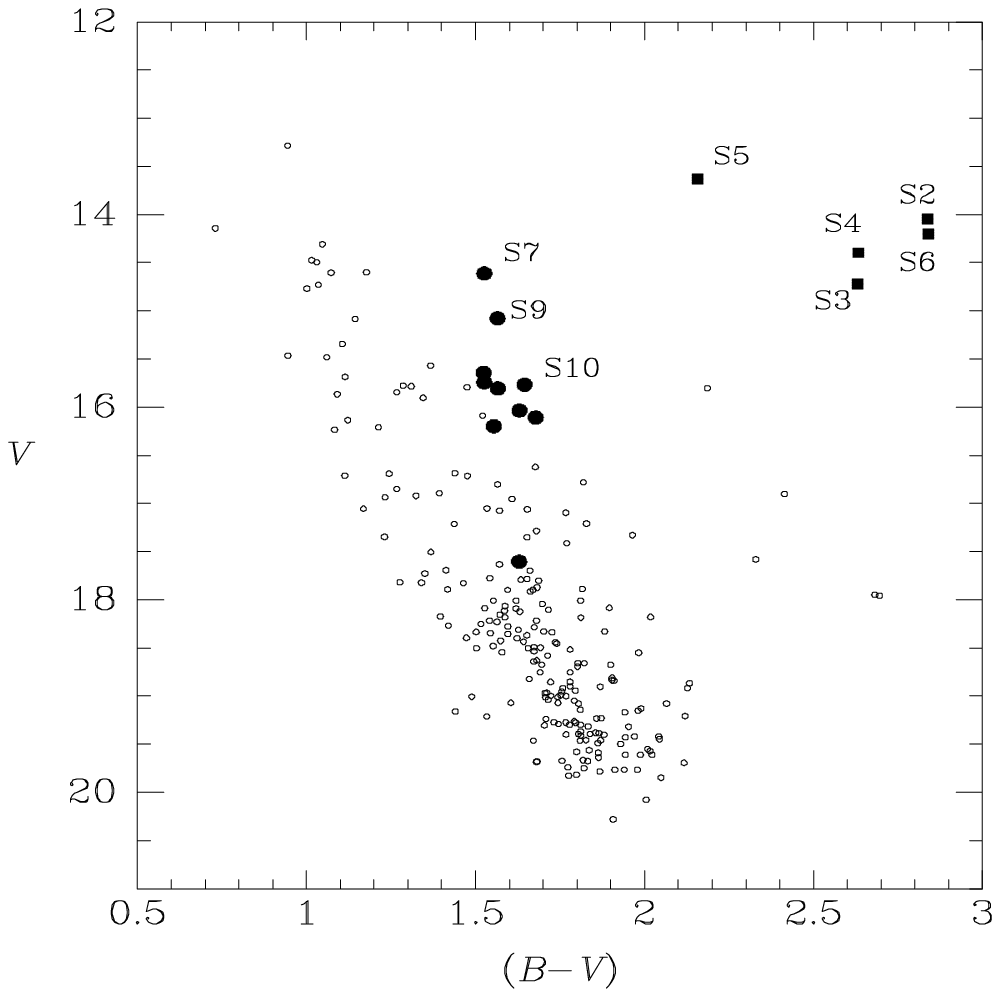}
   \caption{$V/(B-V)$ diagram for all stars in the field of Be~55. Filled circles represent B-type stars observed spectroscopically and filled squares are supergiant stars.\label{rawcmd}}
    \end{figure}

We start the photometric analysis by plotting the $V/(B-V)$ and $V/(U-B)$ diagrams for all stars in the field. In Figure~\ref{rawcmd}, we can observe that the cluster sequence (as defined by the spectroscopic members) is heavily contaminated by what seems to be foreground population.

\subsubsection{The reddening law}
The first step is determining whether the extinction law in the
direction of the field is standard. We start by selecting the B-type stars observed spectroscopically. For them, we calculate the ratio 
$X=E(U-B)/E(B-V)$ using the calibration of intrinsic colors by \citet{fitzgerald70}. For 6 stars, we find rather homogeneous values, with average $X=0.67\pm0.03$, quite close to the standard value $0.72$.
 
We also use the CHORIZOS ($\chi^2$ code
for parametrized modeling and characterization of photometry and
spectroscopy) code developed by \citet{jesus2004}. This code fits
synthetic photometry derived from the 
spectral distribution of a stellar model convolved with 
an extinction law \citep{cardelli} to reproduce the observed magnitudes. We use our $UBV$ photometry and $JHK_{{\rm S}}$ data from 2MASS. This set of photometric values is well suited to find the general shape of the extinction law \citep{jesus2004}. On the other hand, the functional form of the law used by \citet{cardelli}, a 7th degree polynomial, may lead to an artificial bumpiness of the extinction law for high values of $E(B-V)$.

For the stars with spectral types, we use as input for CHORIZOS the $UBVJHK_{{\rm S}}$ photometry and the $T_{{\rm eff}}$ corresponding to the spectral type derived according to the calibrations of \citet{fitzgerald70}. The output of CHORIZOS is the value of $R$ and the color excess 
$E(B-V)$. For all our stars, the preferred value of $R$ is close to
3.1, with little dispersion.

\subsubsection{Membership determination}
Since the extinction is close to standard, we can use the classical $Q$ parameter \citep{johnson52} to estimate the spectral types for all the stars in the field. A first test with the stars with spectral types supports the use of $Q=(U-B)-0.67(B-V)$, as the standard value 0.72 gives too early spectral types when compared to the spectroscopic ones (see below). Even such small difference in the ratio of color excesses, when combined with the high $E(B-V)$, introduces differences of about two subtypes.

Taking into account the photometric spectral types and the position of the stars in the CMDs, we select the likely members of Be~55. We identify 138 members, all with spectral types between b3 and a0. We proceed then to estimate the reddening for all the likely members with CHORIZOS. As input, we use the $UBVJHK_{{\rm S}}$ photometry and the $T_{{\rm eff}}$ corresponding to the photometric spectral type, according to the calibrations of \citet{fitzgerald70}. We find that the average color excess is $E(B-V)=1.85\pm0.16$, where the error represents the dispersion in individual values amongst all 138 members in our photometric field.       

\begin{figure}
   \centering
   \includegraphics[width=\columnwidth]{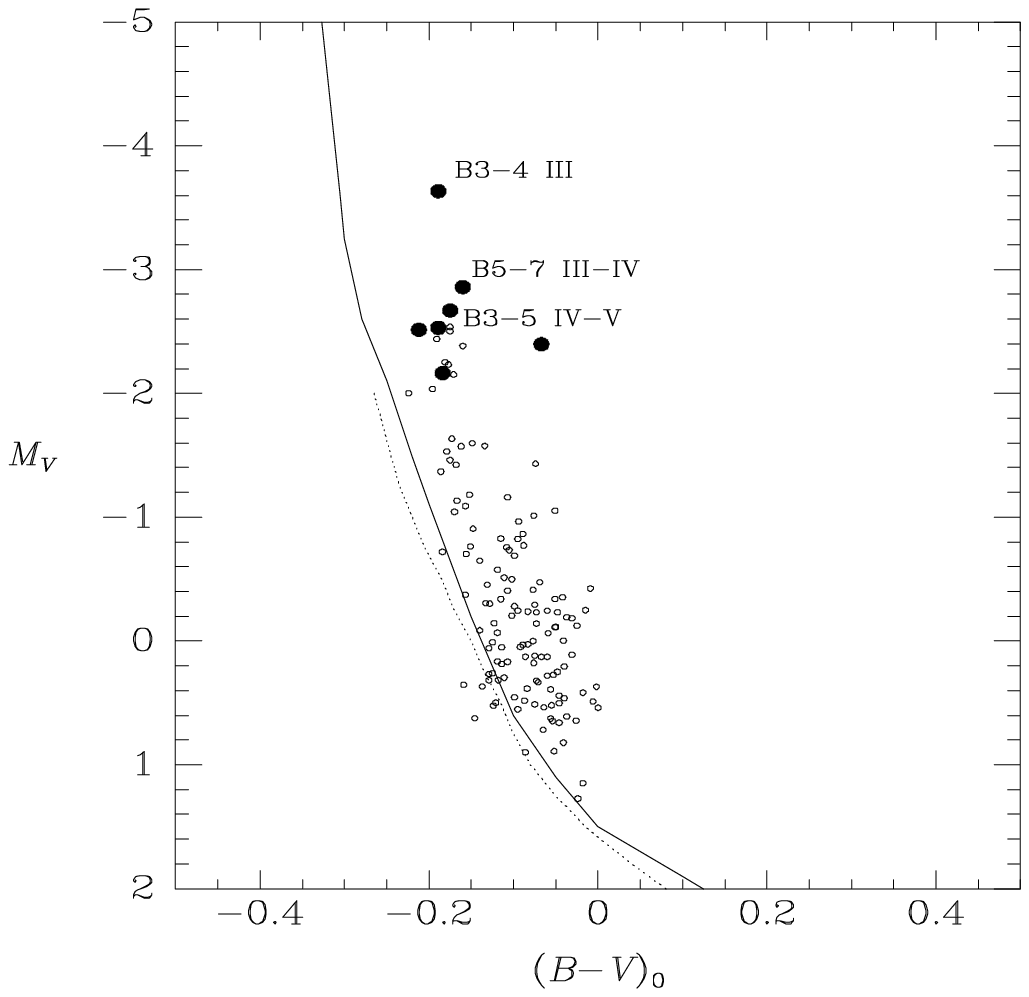}
   \caption{Dereddened $M_{V}/(B-V)$ diagram for likely members of
     Be~55. The dashed line shows the ZAMS from \citet{mermilliod1981}, and the solid line, the ZAMS from \citet{skaler82}.  Filled circles indicate stars with spectra. The spectral types of the brightest members are shown. Note that the Be shell star S9 is not selected as a likely member because of its anomalous colors. \label{clean}}
   
    \end{figure}

In Fig.~\ref{clean}, we plot the dereddened $M_{V}/(B-V)$ diagram for likely members of Be~55. We perform a visual fit to the ZAMS from \citet{mermilliod1981} and from \citet{skaler82}, obtaining a distance modulus $DM=13.0\pm0.3$. In Table~\ref{membersphot}, we display the values of $E(B-V)$, $V_{0}$ and photometric spectral types for all photometric members.

\begin{table*}
\centering
\caption{Photometry and reddening for stars selected as likely members of Berkeley 55\tablenotemark{a}.\label{membersphot}}
\begin{tabular}{lccccc}
  \tableline
  \tableline
Star&$E(B-V)$&$\sigma_{E(B-V)}$&$V_{0}$&$\sigma_{V_{0}}$&Photometric spectral type\\
\tableline
1	&	1.67	&	0.03	&	13.26	&	0.06	&	b8	\\
4	&	2.15	&	0.02	&	13.36	&	0.07	&	b7	\\
5	&	1.76	&	0.02	&	14.27	&	0.11	&	b9	\\
6	&	1.74	&	0.03	&	12.53	&	0.04	&	b8	\\
7	&	1.99	&	0.02	&	13.32	&	0.09	&	b8	\\
8	&	1.89	&	0.02	&	13.51	&	0.07	&	b9	\\
10	&	1.94	&	0.02	&	13.52	&	0.07	&	b8	\\
...&...&...&...&...&...\\
\tableline
\end{tabular}
\tablenotetext{a}{The two known Be stars, S9 (\#199) and S19 (\#213) have been left out, as their intrinsic color cannot be calculated.}
\end{table*}

\subsection{Spectroscopic analysis}

Because of the high reddening, stars in Berkeley~55 are faint in $B$, and intermediate-resolution spectra in the classification region would require long integrations with large apertures. Alternatively, we resort to the far red, where the stars are much brighter and classification criteria exist. 

\subsubsection{Bright red stars}

\begin{figure*}
\resizebox{\textwidth}{!}
   {\includegraphics[angle=-90]{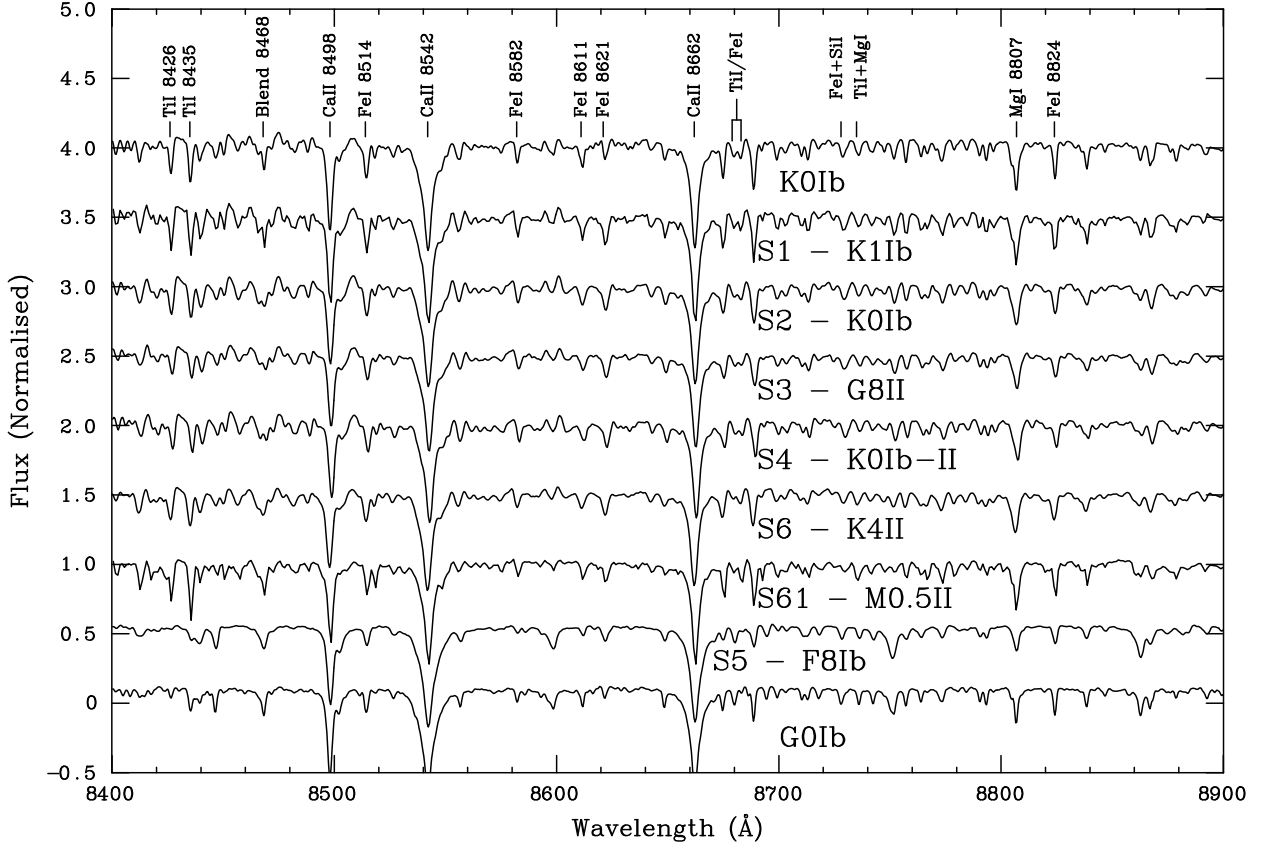}}
  \caption{$z$-band spectra of seven bright red stars in the field of
    Be~55,  together with two reference stars from the Indo-US library \citep{indous}. The main temperature criterion for G--K luminous stars in this range is the Ti\,{\sc i}~8683\AA/Fe\,{\sc i}~8679\AA\ ratio. The strength of the metallic spectrum is the main criterion to determine luminosity at a given spectral type, with the Ti\,{\sc i}/Fe\,{\sc i} blend at 8468\AA\ showing a pronounced dependence. S5 has a much earlier spectral type and the Paschen lines are still seen.  The top spectrum corresponds to HD~207089; the bottom one, to HD~204867 \citep[both classified in][]{keenan}. \label{fig:sgs}}
\end{figure*}

Figure~\ref{fig:sgs} shows the $z$-band spectra of the 7 bright red
stars in the field preselected as candidate members. Six of them have
pretty similar spectra, typical of late-type stars, with S5 appearing clearly earlier.

\begin{table}
 \centering
 \begin{minipage}{\columnwidth}
  \caption{Measured equivalent widths\tablenotemark{1} (in \AA) for the main quantitative luminosity indicators  for red luminous stars in Be~55. \label{caii} }
  \begin{tabular}{lcccc}
  \tableline
   Star   & Ca\,{\sc ii}~8498.0\AA &Ca\,{\sc ii}~8542.1\AA& Ca\,{\sc ii}~8662.1 & Blend~8468\AA    \\
 \tableline
 \tableline
S1  &            3.2  &             5.2&              4.0 & 1.4\\
S2  &            3.0  &             5.3&              4.0 & 1.3\\
S3  &            2.7  &             4.9&              3.6 & 1.0\\
S4  &            3.6  &             5.0&              3.6 & 1.2\\
S5\tablenotemark{2}  &            2.9  &             6.0&              4.9 & $-$\\
S6  &            2.8  &             4.9&              3.5 & 1.2\\
S61 &            2.4  &             4.7&              3.7 & 1.1\\
\tableline
\end{tabular}

\tablenotetext{1}{Because of the
    difficulty in determining the continuum in red stars,
    uncertainties may be estimated at $\pm0.3$\AA\ for each Ca\,{\sc ii} line and $\pm0.1$\AA\ for the 8468\AA\ blend.}
\tablenotetext{2}{Note that the values for S5 must include the blended Paschen lines.}
\end{minipage}
\end{table}

In late-type stars, at a given metallicity, the strength of the metallic spectrum increases with luminosity \citep{ginestet94,carquillat97}. Luminosity criteria at resolutions similar to ours are discussed in \citet{neg11}. In particular, the strength of the Ca\,{\sc ii} triplet is strongly correlated with luminosity \citep{diaz89,zhou91}, though it also depends on metallicity and spectral type. The intensity of the Ca\,{\sc ii} triplet lines (see measurements in Table~\ref{caii}) shows without doubt that all our 7 objects are luminous stars.

The O\,{\sc i}~7774\AA\ triplet is seen in stars earlier than late G. Of all our bright red stars, S5 is the only one displaying this feature. On the other hand, none of our stars displays TiO bands in the spectral range measured, showing that they are earlier than $\sim$M1 \citep[e.g.,][]{neg11}.
For luminous stars in the G5--K5 range, the ratio between the Ti\,{\sc i}~8683\AA\ and the Fe\,{\sc i}~8679\AA\ lines is very sensitive to temperature \citep{carquillat97}. By using this criterion, we find all the stars to lie between G8 and K4, with the exception of S61, which is slightly later, though it is still earlier than M1, as the TiO bandhead at 8660\AA\ is hardly detectable.

 For spectral types G--K , \citet{diaz89} find that stars with a combined equivalent width (EW) for the two strongest Ca\,{\sc ii} lines,  EW(Ca\,{\sc ii}~8542+Ca\,{\sc ii}~8662)$>$9\AA\ are definitely supergiants, independently of metallicity or spectral type, though some supergiants may present slightly lower values of EW.  This has recently been re-assessed for a larger sample \citep{neg11}.
 Attending to all the criteria and to direct comparison to the standards, S1 and S2 fall clearly in the supergiant region, though both have low luminosity. All the other objects lie in the region where both supergiants and bright giants are found. Attending to their morphology, we classify them as bright giants, with S4 closer to being a supergiant than the others. The spectral types finally derived are listed in Table~\ref{spectypes}.

In the case of S5, comparison to MK standards observed at similar resolution \citep[e.g.,][]{cenarro} suggests it is an F8 supergiant. The combined EW of the two strongest
Ca\,{\sc ii} lines is close to $11$\AA, but the Paschen lines are still present at this spectral type and thus Pa~13 and Pa~15 must be blended with Ca\,{\sc ii}~8662\AA\ and Ca\,{\sc ii}~8542\AA, respectively. By measuring the EWs of Pa~12, Pa~14 and Pa~17, we estimate that they contribute at most 1.5\AA\ to the blend. Therefore S5 is a supergiant
according to all calibrations of the Ca\,{\sc ii} triplet \citep{diaz89, mallik97, zhou91}. The calibration of $M_{V}$ against the EW of
the O\,{\sc i}~7774\AA\ triplet \citep{arel03} indicates for this
object $M_{V}=-3.7$, with a large uncertainty. This is consistent with the expectations for a such a supergiant, especially if it is moving redwards for the first time. In view of this, we accept the morphological classification F8\,Ib. This object must thus lie on the instability strip.

\begin{table}
\centering
 \begin{minipage}{\columnwidth}
 \caption{Derived parameters for stars with spectroscopy.\label{spectypes}} 
\begin{tabular}{lcccccccc}
\tableline
\tableline
ID&Spectral &$K_{{\rm S}}$\tablenotemark{1}&$(J-K_{{\rm S}})$&$(J-K)_{0}$&$E(J-K_{{\rm S}})$\\
&type\tablenotemark{2,3} & &&&\\
\tableline
S1	&K1\,Ib &6.25	&1.66&	0.68&	0.98\\
S2	&K0\,Ib	&6.65	&1.59&	0.63&	0.96\\
S3	&G8\,II	&7.80	&1.47&	0.56&	0.91\\
S4	&K0\,Ib-II&7.33 &1.47&	0.63&	0.84\\
S5	&F8\,Ib	&7.71	&1.21&	0.35&	0.86\\
S6      &K4\,II &6.48   &1.70&  0.90&   0.80\\
S61	&M0.5\,II&6.15&1.55&	1.04&	0.51\\
\tableline
S7&B3--4\,III (b4)&10.47&0.68&$-0.15$&0.83\\
S8&B6--8\,III (b5)&11.52&0.73&$-0.08$&0.81\\
S9&B3--4\,IIIshell&10.63&0.73&$-$&$-$\\
S10&B5--7\,III--IV (b5)&11.20&0.78&$-0.10$&0.88\\
S12&B3--5\,IV--V (b4)&11.50&0.74&$-0.13$&0.87\\
S13&B3--5\,IV--V (b5)&11.62&0.74&$-0.13$&0.87\\
S15&B4--5\,IV--V (b5)&11.95&0.75&$-0.12$&0.87\\
S16&B5--7\,III--IV (b5)&11.56&0.67&$-0.12$&0.79\\
\tableline
\end{tabular}
\tablenotetext{1}{The typical uncertainty of the photometry is 0.02~mag in $K_{{\rm S}}$ and 0.03~mag in $(J-K_{{\rm S}})$.}
\tablenotetext{2}{The whole range of spectral types and luminosity classes compatible with the spectral information available is shown.}
\tablenotetext{3}{Photometric spectral types are given in parentheses.}
\end{minipage}
\end{table}

\subsubsection{Blue stars}

\begin{figure}
\resizebox{\columnwidth}{!}
   {\includegraphics{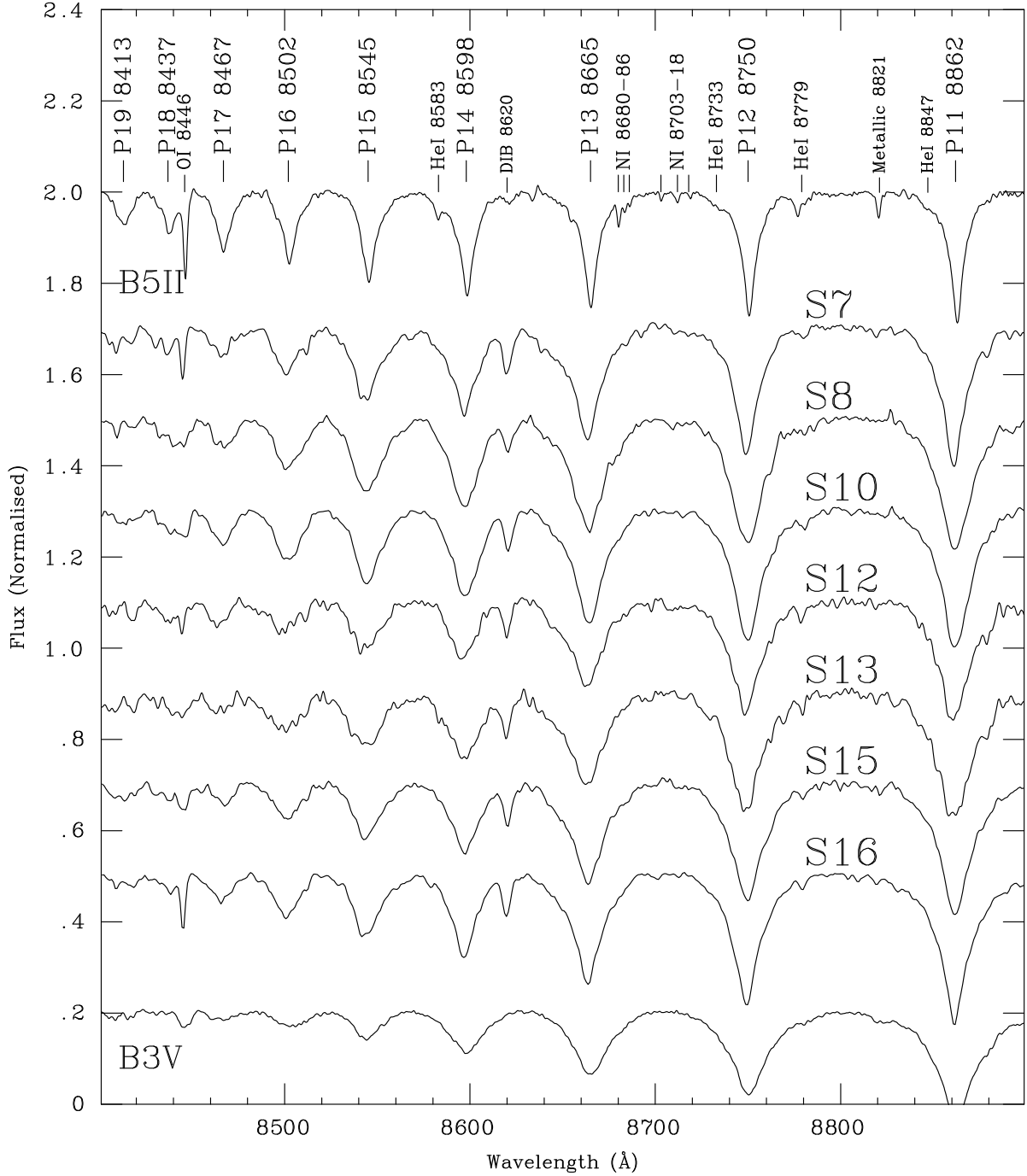}}
  \caption{$z$-band spectra of a sample of intrinsically blue stars in
    Be~55, together with two reference stars from the Indo-US library \citep{indous}. The top spectrum is HD~164353. The bottom spectrum is HD~120315.\label{fig:bstars}}
\end{figure}

\begin{figure}
\resizebox{\columnwidth}{!}
   {\includegraphics[angle=-90]{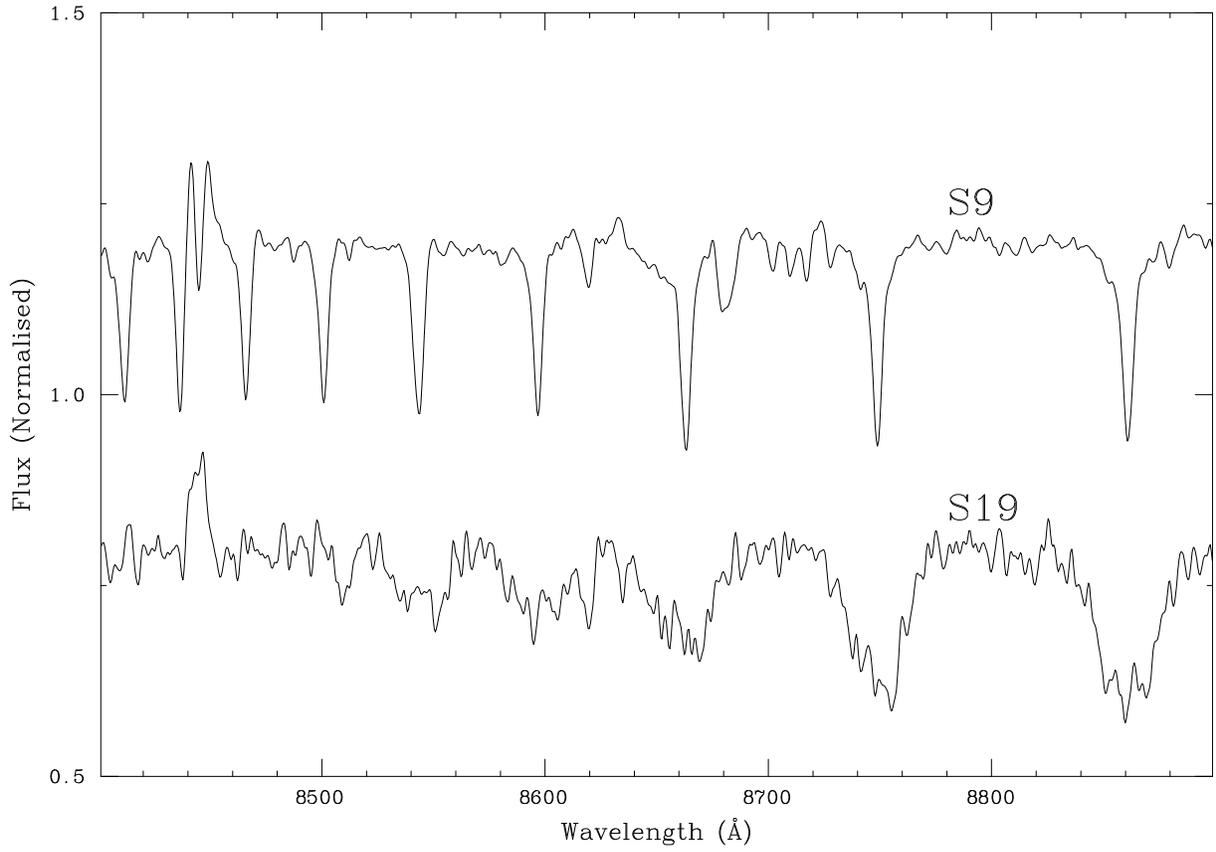}}
  \caption{$z$-band spectra of two Be stars in Be~55. S9 is a shell
    star and the second brightest blue star in the cluster.\label{fig:bestars}}
\end{figure}

Figure~\ref{fig:bstars} shows spectra of stars
with intrinsically blue spectra, while Fig.~\ref{fig:bestars} shows the spectra of two stars with emission lines. The brightest early-type objects are S7 and S9. The $z$-band is not rich in classification features for B-type
stars \citep{and95,munarit99}. Moderately accurate spectral types can be achieved for supergiants, but stars close to the main sequence are poorer in spectral features and can only receive approximate classifications \citep{negwd1}. Apart from the strength and width of Paschen lines, the
main criterion is the presence of He\,{\sc i} lines, which are only
moderately strong in supergiants and are hardly seen in mid and late
main-sequence B-type stars. All the spectra displayed are clearly
dominated by broad Paschen lines and are therefore B-type stars of low or moderate luminosity. The only exception is S9, which looks like an A-type
supergiant. However, the presence of O\,{\sc i}~8448\AA\ in emission
betrays this as a Be star. Some Be shell stars have absorption
spectra (due to the circumstellar disk seen close to edge on) that strongly resemble A-type supergiants \citep{and88}.

Star S7 has very well defined Paschen lines and resembles rather
closely the B5\,II star in Fig.~\ref{fig:bstars}, though it has lower luminosity. The narrow
O\,{\sc i}~8448\AA\ indicates a slow rotator and most likely a
giant. This is confirmed by the sharpness of the Paschen lines
(compare to the B3\,V star). We estimate a spectral type
B3--4\,III. S16 is very similar, though it rotates faster or has a
lower luminosity. S10 is a fast rotator, but is likely a
giant as well, as a weak He\,{\sc i}~8779\AA\ seems to be present. S12, S13
and S15 are closer to the main sequence. The fact that O\,{\sc i}~8448\AA\ is
weak compared to Paschen~18 places them close to B3. The spectrum of S19 (Fig.~\ref{fig:bestars}) is very noisy,
but O{\sc i}~8448\AA\ is clearly in emission. Therefore it is a Be
star. According to \citet{and88}, late Be stars do not show this line
in emission, but only filled-in. Therefore this star is B5\,Ve or
earlier. Based on the same criterion and its brightness, S9 is likely
to be a shell star of spectral type similar to S7. Finally, S8 looks
later than all the others, though its reddening is similar to those of
the other B stars. A summary of
estimated spectral types is given in Table~\ref{spectypes}.

\citet{caron03} developed a method to classify B-type stars by comparing the EWs of the Paschen lines. Application of those criteria suggest slightly later spectral types for our stars, though not all criteria suggest the same result (something expected, as they are only approximate). Following them, S7 would be $\sim$B5\,III, S8 and S10 would be $\sim$B8\,III, and the rest of the stars would be $\sim$B5\,V.  These small differences are in all likelihood due to the continuum determination, which is somewhat subjective, and in the case of the Paschen lines (with their very broad wings) may depend to some degree on SNR and resolution.


\subsection{Infrared extinction}

From the spectral types derived, we can calculate individual infrared reddenings. For the red (super)giants, we use the calibration of \citet{straizys09}, which uses 2MASS colors. For the blue stars, we use the same calibration, assuming that giants have the same colors as main-sequence stars. For the F supergiant, we use the value from \citet{koorn83}. The values derived are shown in Table~\ref{spectypes}.  The reddening seems to be moderately variable across the face of the cluster. The values for the supergiants show a higher dispersion than those of the blue stars, but the average reddening for the 6 red (super)giants (excluding the outlier S61) is $E(J-K_{S})=0.89\pm0.07$ (the error
is the standard deviation), while the average for 7 B-type stars without emission lines is $E(J-K_{S})=0.85\pm0.04$, fully compatible within their errors. Using the calibrations of \citet{winkler} or \citet{wegner} for the blue stars produces slightly different individual values, but compatible averages.

Comparison of the $E(B-V)$ (Table~\ref{membersphot}) and $E(J-K_{{\rm S}})$ excesses for these objects (Table~\ref{spectypes}) confirms that the extinction law in this direction is close to the $R=3.1$ galactic average, for which $E(B-V)\approx2\times E(J-K)$. It is also interesting to note that the B-type stars spectroscopically observed consistently have $E(B-V)$ below the cluster average. This suggest that some of the faint late-B stars taken as photometric members could be background objects.

\subsection{Cluster age}
\label{sec:hr}

The dereddened CMD for Be~55 (Fig.~\ref{clean}) shows that stars brighter than $M_{V}\approx-1$ start deviating from the ZAMS. This absolute magnitude corresponds to spectral type B4\,V \citep[e.g.,][]{turner80}, in good agreement with the determination of spectral types, which indicate that the brightest B-type stars are B3--6\,III-IV objects. With such a turn-off, the cluster is expected to have an age $\ga40$~Myr. We can improve on this value by fitting isochrones in the different observational CMDs to the position of the red stars and the turn-off.


\begin{figure}
   \resizebox{\columnwidth}{!}
   {\includegraphics{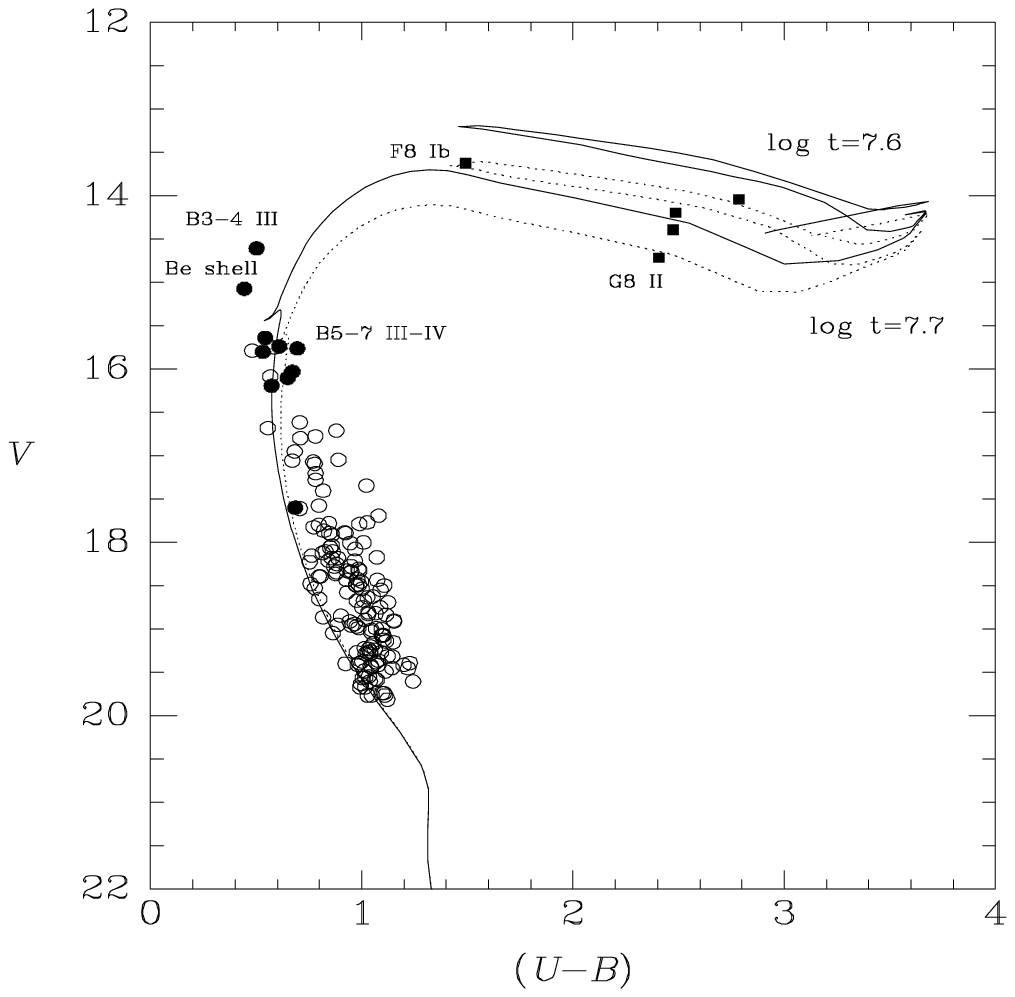}}
   \caption{$V/(U-B)$ diagram for likely members of Berkeley~55. Filled circles are confirmed spectroscopic members, while open circles  are likely (photometric) members. Two isochrones of \citet{marigo} are shown, both reddened by $E(B-V)=1.85$ and displaced to $DM=13.0$. The isochrones have been reddened taking into account color effects, following the procedure described in \citet{girardi08}.\label{Figubv}}
   
    \end{figure}


A fundamental issue is the correct determination of color excesses at such high reddenings. Figure~\ref{Figubv} shows the observational $V/(U-B)$ diagram for the cluster together with isochrones for $\log t=7.6$ (40~Myr) and  $\log t=7.7$ (50~Myr) from \citet{marigo}. Both isochrones fit well the position of the main sequence and the turn-off. S7 appears as a mild blue straggler, which is consistent with its spectral type (a B3--4\,III giant when the turn-off is at B4\,V). S9 may also be a blue straggler, but shell stars are known to have anomalous $(U-B)$ colors. The isochrones have been reddened following the procedure described in \citet{girardi08}, which takes into account color effects for late-type stars. In spite of this, the position of the K (super)giants does not fit the red clump very well for any of the two isochrones. This is likely due to the difficulty in reproducing the extinction law with the polynomial approximation of \citet{cardelli}.

\begin{figure}
   \resizebox{\columnwidth}{!}
   {\includegraphics{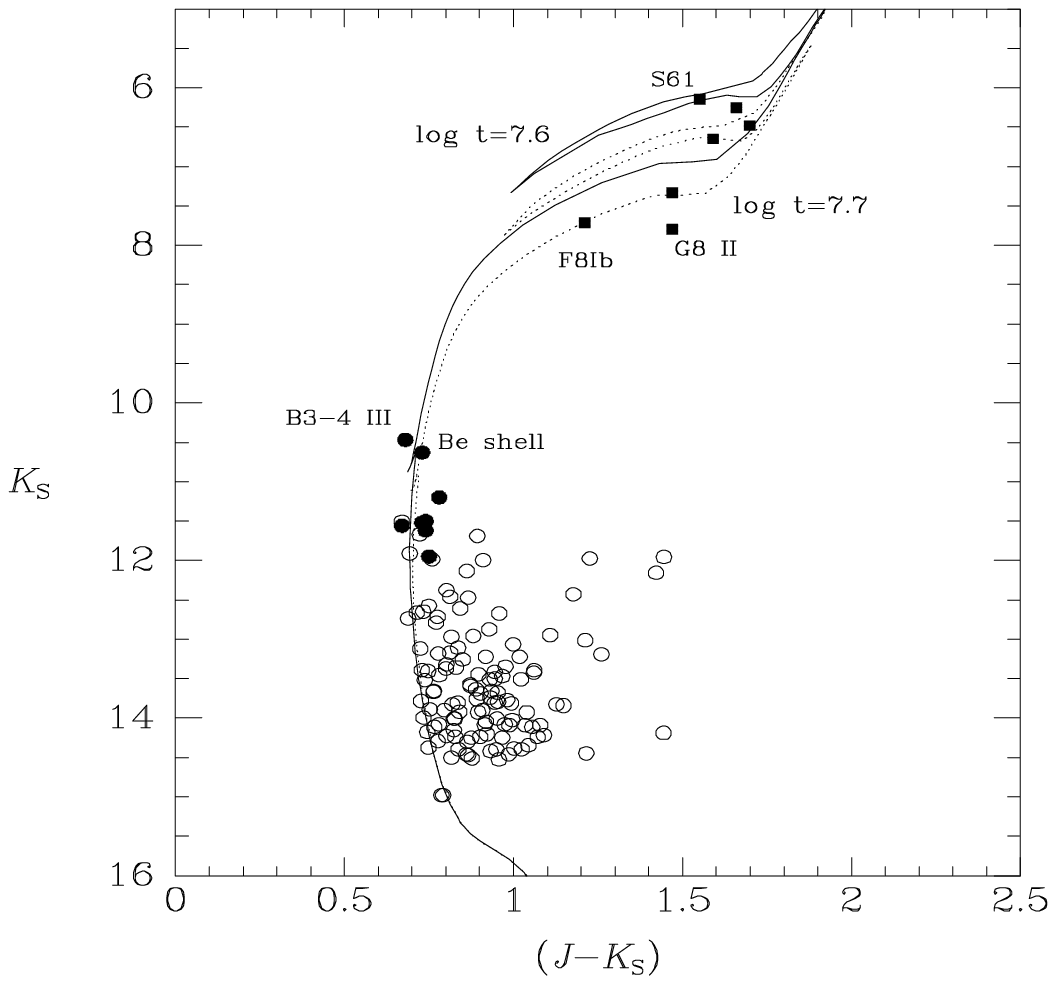}}
   \caption{$K_{{\rm S}}/(J-K_{{\rm S}})$ diagram for for likely members of Berkeley~55. Filled circles are confirmed spectroscopic members, while open circles are likely (photometric) members. Two isochrones of \citet{marigo} are shown, both reddened by $E(J-K_{{\rm S}})=0.85$ and displaced to $DM=13.0$. \label{Figkj}}
  
    \end{figure}

Figure~\ref{Figkj} shows the $K_{{\rm S}}/(J-K_{{\rm S}})$ CMD for cluster members, with photometry from 2MASS. Isochrones from \citet{marigo} for the same ages are shown, reddened by $E(J-K_{{\rm S}})=0.85$, the value found for the blue spectroscopic members. All members selected from their $UBV$ colors fall to the right of the isochrones, confirming that this excess corresponds to the less reddened members. Some likely members seem to have much higher color excesses, suggesting that they are either Be stars or background B-type stars (though this is not generally supported by their photometric spectral types). In this CMD, the K (super)giants occupy positions in very good agreement with the 50~Myr isochrone.


The bright star S61, situated $\sim4\farcm7$ from the cluster core falls together with the red (super)giants in this CMD. However, its position does not fit well with its spectral type, later than all the other evolved stars. As seen in Table~\ref{spectypes}, its reddening is rather lower than that of confirmed members. Its $(B-V)$ magnitude, taken from \citet{macie07}, is compatible with the low reddening. Therefore this object could be a foreground star as well as a cluster member with much lower reddening. This should be checked via radial velocity measurements\footnote{Even though our spectra were not taken to measure radial velocities, direct cross-correlation between the spectra of S1 and S61 (the two objects observed with the same configuration) suggests that their radial velocities are very similar}.

\section{Discussion} \label{discussion}

We have presented spectroscopy of a sizable sample of stars in the open cluster Berkeley~55, revealing a population of 7 low-luminosity supergiants or bright giants (one of them, not a certain member), 4 mid-B giants and a main sequence starting around B4\,V. 

\subsection{Cluster parameters}

Only two previous studies of the cluster have been published. \citet{macie07} derive an age of $\sim315$~Myr. \citet{tadross} used their result to calibrate his fitting procedure and therefore derives a similar age of 300~Myr, from a fit to the 2MASS data alone.  This age is completely incompatible with the observed population, as stars earlier than B5 are only seen in clusters younger than 100~Myr. 
Clusters with $\sim100$~Myr have stars of spectral type $\sim$B6 around the main-sequence turn-off (e.g., NGC~7790, \citealt{matthews95}; Berkeley~58, \citealt{turner08}).

The difference between our age determination and that of \citet{macie07} is almost certainly due to the identification of the evolved population. The decontaminated $V/(B-V)$ in \citet{macie07} does not show a clump of red stars, as in our Fig.~\ref{rawcmd}. Some of the (super)giants seem to be blended in their images and do not appear in their original photometry. The others are not conspicuous when a wide field is considered, because of contamination by field interlopers with similar $(B-V)$ color. The stars forming this clump, however, cannot be overlooked when the infrared data are considered. S1 is the brightest star in $K_{{\rm S}}$ within $10\arcmin$ of the cluster center, and there are only 5 field stars with $K_{{\rm S}}$ magnitudes comparable to the 7 red (super)giants. The 7 candidate members form a clearly separated clump in the $(J-K_{{\rm S}})$/$K_{{\rm S}}$, $H-K_{{\rm S}}$/$K_{{\rm S}}$ and $Q$/$K_{{\rm S}}$ diagrams, and have similar (and unusual) spectral types. Though accurate radial velocities will be needed to confirm individual memberships, there can be no doubt that they represent the population of evolved stars in the cluster. 
The small systematic offset in $(B-V)$ between our photometry and that of \citet{macie07} is unlikely to have biased their age determination. It is simply very difficult to obtain an accurate fit when photometry in only two bands is available.

 At Galactic longitude $\ell=93\fdg0$, a distance $\sim4$~kpc (corresponding to our $DM=13.0$) seems compatible  with tracers of the Perseus arm, though some Galactic models suggest a higher distance for this feature. Spiral tracers in this direction are scarce. NGC~7067 ($\ell=91\fdg2$) has distance estimates of 3.6~kpc \citep{ys04} and 4.4~kpc \citep{hassan73}. NGC~7128 ($\ell=97\fdg3$), with $DM=13.0\pm0.2$ (4~kpc), is another likely tracer at some distance from Be~55 \citep{balog}.

As the age of Berkeley~55 is $\sim50$~Myr, the red stars in the cluster should have initial masses $M_{*}\ga7.0\,M_{\odot}$ \citep{marigo}, or slightly higher if they started their lives rotating quickly \citep[cf.][]{mm00}. Such stars define the boundary between intermediate-mass and massive stars, and therefore can provide important constraints for stellar evolution models.  Milky Way clusters with similar ages and rich populations of RSGs (or bright giants) are relatively scarce. NGC~6649 ($\sim50$~Myr; \citealt{turner81}) with one Cepheid and four red (super)giants is one of the few examples. NGC~6664 (also $\sim$50~Myr; \citealt{schmidt82}) has a similar population, but is less well studied. As a consequence, the sample of Galactic RSGs studied by \citet{levesque05} contains almost no stars in the $6\,M_{\odot}<M_{*}<10\,M_{\odot}$ range.  The presence of 7 candidate evolved members makes Berkeley~55 a very suitable target to study the physical effects operating during the He-core burning phase of intermediate-mass stars. 

\subsection{The lowest mass for a supergiant}

As mentioned in the introduction, the separation between red supergiants and red bright giants is morphological, and does not coincide with the separation between high-mass and low-mass stars. Of course, making such a distinction necessarily implies dividing a continuum into artificial boxes\footnote{It should be noted that many authors classically count luminosity class II objects as supergiants}. In spite of that, defining an age limit after which clusters do not contain RSGs seems interesting, as new clusters are being discovered where high extinction renders only the bright red stars accessible for spectroscopy \citep[e.g.,][]{neg10,davies11}.

Looking at the lists of yellow evolved stars in clusters from \citet{harris76} or \citet{sowella,sowellb}, it is difficult to set an age limit after which Ib supergiants are not seen. A few objects have been classified as K\,Ib supergiants in clusters $\ga 100$~Myr old, but in all cases there are alternative lower-luminosity classifications based on more recent data (for example, star 164 in NGC~129, classified K2.5\,II-III by \citealt{keenan}\footnote{This star seems not to be a member, based on its radial velocity \citep{mermi08}, though it was considered a member for a long time \citep[see references in][]{turner92}.}; or star 110 in NGC~2516, classified M1.5\,IIa by \citealt{kp85}). Most classical Cepheids appear as F/G Ib supergiants.  Many clusters containing Cepheids are sparsely populated and have the Cepheid as only supergiant \citep[e.g.,][]{turner96}. In clusters with larger populations, the ratio of F/G supergiants to later spectral types is very variable. For example, NGC~7790 contains three F-type Cepheids, but no red (super)giants, while NGC~6067 (with two Cepheid members) contains a large number of red giants and supergiants \citep{mer87}. 

 If we consider a number of clusters for which accurate ages in the range of interest have been derived through careful photometric analysis or equivalent methods, we find by decreasing age, that NGC~2516 ($\sim140$~Myr), which has no Cepheids, contains four bright red giants, with luminosity classes II or III \citep{kp85,gl00}. IC~4725, which is $\sim90$~Myr old, contains one Cepheid supergiant and two class II red giants \citep{kp85,mer87}. Berkeley~82, with an age $\sim70$~Myr contains two yellow bright stars with spectral types G2\,II/Ib and K2\,II \citep{forbes86}, while Harvard~20 contains a G5\,Ib supergiant at $\sim60$~Myr \citep{turner80}. On the other hand, NGC~6664, with an age $\sim50$~Myr, contains several objects classified as bright giants by \citet{harris76}. Finally, NGC~6067 contains red stars with luminosity classes II and Ib, leading \citet{mer87} to doubt its estimated age $\ga100$~Myr, which is also in contradiction with the observed period of its fundamental-mode Cepheid member V340~Nor.  

In view of this, it seems that red stars close to the clump position are classified as K supergiants in clusters up to 50--60~Myr, while some stars are classified as such in somewhat older clusters. No cluster older than $\sim100$~Myr seems to contain a bona-fide K supergiant. Stars with F and G spectral types, such as Cepheids, are classified as supergiants at older ages, e.g., SZ Tau (F5-9\,Ib) in NGC~1647 \citep[$\sim150$~Myr;][]{turner92}. 

The classifications derived for stars in Be~55 seem to confirm these trends, with stars on both sides of the divide at an age $\approx50$~Myr.

\acknowledgments

This research is partially supported by the Spanish Ministerio de
Ciencia e Innovaci\'on (MICINN) under
grants AYA2008-06166-C03-03, AYA2010-21697-C05-05 and CSD2006-70, and by the Generalitat Valenciana (ACOMP/2010/259).

 The WHT is operated on the island of La
Palma by the Isaac Newton Group in the Spanish Observatorio del Roque
de Los Muchachos of the Instituto de Astrof\'{\i}sica de Canarias. The
July 2007 observations were taken as part of the service program
(program SW2007A63).

Based on observations made with the Nordic Optical Telescope, operated
on the island of La Palma jointly by Denmark, Finland, Iceland,
Norway, and Sweden, in the Spanish Observatorio del Roque de los
Muchachos of the Instituto de Astrof\'{\i}sica de Canarias.  

Some of the data presented here have been taken using ALFOSC, which is owned by the Instituto de Astrofisica de Andalucia (IAA) and operated at the Nordic Optical Telescope under agreement between IAA and the NBIfAFG of the Astronomical Observatory of Copenhagen

This research has made use of the Simbad database, operated at CDS,
Strasbourg (France) and of the WEBDA database, operated at the
Institute for Astronomy of the University of Vienna. This publication
makes use of data products from 
the Two Micron All Sky Survey, which is a joint project of the University of
Massachusetts and the Infrared Processing and Analysis
Center/California Institute of Technology, funded by the National
Aeronautics and Space Administration and the National Science
Foundation.



{\it Facilities:} \facility{WHT(ISIS), NOT(ALFOSC)}


\clearpage

\begin{table*}
\caption{$(X,Y)$ position on the map of all stars with $UBV$ photometry, together with their 2MASS identification and their coordinates.\label{allpos}}
\begin{tabular}{lccccc}
  \tableline
  \tableline
Star&$X$ (Pixels)&$Y$ (Pixels)&RA (J2000)&Dec (J2000)& Name (2MASS)\\
\tableline
 1 & 15.28 & 1446.924 & 319.327021 & 51.790562 & 21171848+5147260 \\
 2 & 21.685 & 1886.522 & 319.326885 & 51.813904 & 21171845+5148500 \\
 3 & 30.195 & 943.208 & 319.3254 & 51.763935 & 21171809+5145501 \\
 4 & 31.361 & 531.363 & 319.325026 & 51.742191 & 21171800+5144318 \\
 5 & 33.537 & 991.763 & 319.325118 & 51.766521 & 21171802+5145594 \\
 6 & 43.785 & 1622.479 & 319.324686 & 51.799866 & 21171792+5147595 \\
 ...&...&...&...&...&...\\
\tableline
\end{tabular}
\end{table*}

\begin{table}
\caption{Photometry for stars in Berkeley~55.\label{allphot}}
\begin{tabular}{lccccccc}
  \tableline
  \tableline
Star&$V$&$\sigma_{V}$&$(B-V)$&$\sigma_{(B-V)}$&$(U-B)$&$\sigma_{(U-B)}$&$n$\\
\tableline
1	&	18.218	&	0.038	&	1.541	&	0.028	&	0.842	&	0.013	&	2	\\
2	&	18.086	&	0.040	&	1.527	&	0.030	&	1.433	&	0.075	&	2	\\
3	&	19.400	&	0.024	&	1.768	&	0.030	&	1.361	&	0.080	&	1	\\
4	&	19.612	&	0.000	&	1.988	&	0.012	&	1.037	&	0.090	&	2	\\
5	&	19.274	&	0.011	&	1.732	&	0.034	&	0.976	&	0.024	&	2	\\
6	&	17.898	&	0.037	&	1.670	&	0.024	&	0.846	&	0.019	&	2	\\
7	&	19.769	&	0.029	&	1.912	&	0.040	&	1.027	&	0.091	&	1	\\
8	&	19.145	&	0.006	&	1.810	&	0.033	&	1.114	&	0.060	&	2	\\
9	&	16.891	&	0.021	&	1.393	&	0.041	&	0.580	&	0.011	&	4	\\
...&...&...&...&...&...&...&...\\
\tableline
\end{tabular}
\end{table}





\begin{thebibliography}{}
\bibitem[\protect\citeauthoryear{Andrillat, Jaschek \& Jaschek}{Andrillat et al.}{1988}]{and88} Andrillat, Y., Jaschek, M., \& Jaschek, C. 1988, A\&AS, 72, 129 

\bibitem[\protect\citeauthoryear{Andrillat, Jaschek \& Jaschek}{Andrillat et al.}{1995}]{and95} Andrillat, Y., Jaschek, C., \& Jaschek, M. 1995, A\&AS, 112, 475 

\bibitem[\protect\citeauthoryear{Arellano Ferro, Giridhar \& Rojo
    Arellano}{Arellano Ferro et al.}{2003}]{arel03} Arellano Ferro, A., Giridhar, S., \& Rojo Arellano, E. 2003, RMxAA, 39, 3 

 \bibitem[\protect\citeauthoryear{Balog et al.}{2001}]{balog} Balog, Z., Delgado, A. J., Moitinho, A., F\"ur\'esz, G., Kasz\'as, G., Vink\'o, J., \& Alfaro, E. J. 2001, MNRAS, 323, 872


\bibitem[\protect\citeauthoryear{Cardelli, Clayton \& Mathis}{Cardelli et al.}{1988}]{cardelli} Cardelli, J.A., Clayton, G.C., \& Mathis, J.S. 1988, ApJL, 329, 33

\bibitem[\protect\citeauthoryear{Caron et al.}{2003}]{caron03} Caron, G., Moffat, A.F.J., St-Louis, N., Wade, G.A., \& Lester, J.B. 2003, AJ, 126, 1415


\bibitem[\protect\citeauthoryear{Carquillat et al.}{1997}]{carquillat97} Carquillat, J.M., Jaschek, C., Jaschek, M., \& Ginestet, N. 1997, A\&AS, 123, 5

\bibitem[\protect\citeauthoryear{Cenarro et al.}{2001}]{cenarro} Cenarro, A. J., Cardiel, N., Gorgas, J., Peletier, R.F., Vazdekis, A., \& Prada, F. 2001, MNRAS, 326, 959

\bibitem[\protect\citeauthoryear{Chiosi, Bertelli \& Bressan}{Chiosi et al.}{1992}]{chiosi} Chiosi, C., Bertelli, G., \& Bressan, A. 1992, ARA\&A 30, 235

\bibitem[\protect\citeauthoryear{Comer\'on \& Pasquali}{2005}]{cp05}Comer\'on, F., \& Pasquali, A. 2005, A\&A, 430, 541

\bibitem[\protect\citeauthoryear{Davies et al.}{2011}]{davies11} Davies, B., Bastian, N., Gieles, M., 
Seth, A.C., Mengel, S., \& Konstantopoulus, S.
2011, MNRAS, 411, 1386

\bibitem[\protect\citeauthoryear{D\'{\i}az, Terlevich \& Terlevich}{D\'{\i}az et al.}{1989}]{diaz89} D\'{\i}az, A.I., Terlevich, E., \& Terlevich, R., 1989, MNRAS, 239, 325

\bibitem[Draper et al.(2000)]{Draper2000} Draper, P.W., Taylor, M., \& Allan, A. 2000, Starlink User Note 139.12, R.A.L.

\bibitem[\protect\citeauthoryear{FitzGerald}{1970}]{fitzgerald70} Fitzgerald, M.P. 1970, A\&A, 4, 234 

\bibitem[\protect\citeauthoryear{Forbes}{1986}]{forbes86} Forbes, D. 1986, PASP, 98, 218

\bibitem[\protect\citeauthoryear{Ginestet et al.}{1994}]{ginestet94}
Ginestet, N., Carquillat, J.M., Jaschek, M., \& Jaschek, C. 1994, A\&AS, 108, 359

\bibitem[\protect\citeauthoryear{Girardi et al.}{2008}]{girardi08}
Girardi, L., 
et al. 2008, PASP, 120, 583

\bibitem[\protect\citeauthoryear{Gonz\'alez \& Lapasset}{2000}]{gl00} Gonz\'alez, J.F., \& Lapasset, E. 2000, AJ, 119, 2296

\bibitem[\protect\citeauthoryear{Harris}{1976}]{harris76} Harris, G.L.H. 1976, ApJS, 30, 451

\bibitem[Hassan(1973)]{hassan73} Hassan, S.M. 1973, \aaps, 9, 261

\bibitem[Howarth et al.(1998)]{Howarth1998} Howarth, I., Murray, J., Mills, D., \& Berry, D.S. 1998, Starlink User Note 50.21, R.A.L.

\bibitem[\protect\citeauthoryear{Jeffries}{1997}]{jeff97} Jeffries, R.D. 1997, MNRAS, 288, 585

\bibitem[\protect\citeauthoryear{Johnson \& Morgan}{1952}]{johnson52} Johnson, H.L., \& Morgan, W.W. 1952, ApJ, 117, 313

\bibitem[\protect\citeauthoryear{Keenan \& Pitts}{1985}]{kp85} Keenan, P.C., \& Pitts, R.E. 1985, PASP, 97, 297

\bibitem[\protect\citeauthoryear{Keenan \& McNeil}{1989}]{keenan} 
Keenan, P.C., \& McNeil, R.C. 1989, ApJS, 71, 245

\bibitem[Koornneef(1983)]{koorn83} Koornneef, J. 1983, \aap, 128, 84


\bibitem[\protect\citeauthoryear{Landolt}{1992}]{landolt92} Landolt, A.U. 1992, AJ, 104, 340

\bibitem[\protect\citeauthoryear{Levesque et al.}{2005}]{levesque05} Levesque, E.M., Massey, P., Olsen, K.A.G., Plez, B., Josselin, E., Maeder, A., \& Meynet, G.
2005, ApJ, 628, 973 

\bibitem[\protect\citeauthoryear{Maciejewski \& Niedzielski}{2007}]{macie07} Maciejewski, G. \& Niedzielski, A. 2007, A\&A, 467, 1065

\bibitem[\protect\citeauthoryear{Ma\'{\i}z-Apell\'aniz}{2004}]{jesus2004} Ma\'{\i}z-Apell\'aniz, J. 2004, PASP, 116, 859

\bibitem[\protect\citeauthoryear{Mallik}{1997}]{mallik97} Mallik, S.V., 1997, A\&AS, 124, 359

\bibitem[\protect\citeauthoryear{Marigo et al.}{2008}]{marigo}Marigo, P., Girardi, L., Bressan, A., Groenewegen, M.A.T., Silva, L., \& Granato, G.L.
2008, A\&A, 482, 883

\bibitem[\protect\citeauthoryear{Matthews et al.}{1995}]{matthews95} Matthews, J.M., Gieren, W.P., Mermilliod, J.-C., \& Welch, D.L. 1995, AJ, 110, 2280

\bibitem[\protect\citeauthoryear{Mermilliod}{1981}]{mermilliod1981} Mermilliod, J.-C. 1981, A\&A, 97, 235

\bibitem[\protect\citeauthoryear{Mermilliod}{1987}]{mer87} Mermilliod, J.-C. 1987, A\&AS, 70, 389

\bibitem[\protect\citeauthoryear{Mermilliod, Mayor \& Udry}{Memilliod et al.}{2008}]{mermi08} Mermilliod, J.-C., Mayor, M., \& Udry, S. 2008, A\&A, 485, 303

\bibitem[\protect\citeauthoryear{Meynet \& Maeder}{Meynet \& Maeder}{2000}]{mm00}Meynet, G., \& Maeder, A. 2000, A\&A, 361, 101

\bibitem[\protect\citeauthoryear{Mowlavi \& Forestini}{1994}]{mf94} Mowlavi, N., \& Forestini, M. 1994, A\&A, 282, 843

\bibitem[\protect\citeauthoryear{Munari \& Tomasella}{Munari \& Tomasella}{1999}]{munarit99} Munari, U., \& Tomasella, L. 1999, A\&AS, 137, 521

\bibitem[\protect\citeauthoryear{Negueruela \& Schurch}{2007}]{ns07} Negueruela, I., \& Schurch, M.P.E. 2007, A\&A, 461, 431

\bibitem[\protect\citeauthoryear{Negueruela et al.}{2007}]{neg07} Negueruela, I., Marco, A., Israel, G.L., \& Bernabeu, G. 2007, \aap, 471, 485

\bibitem[\protect\citeauthoryear{Negueruela, Clark \& Ritchie}{Negueruela et al.}{2010a}]{negwd1} Negueruela, I., Clark, J.S., \& Ritchie, B.W. 2010a, A\&A, 516, A78

\bibitem[\protect\citeauthoryear{Negueruela et al.}{2010b}]{neg10} Negueruela, I., Gonz\'alez-Fern\'andez, C., Marco, A., Clark, J.S., \&  Mart\'{\i}nez-N\'u\~{n}ez, S. 2010b, A\&A, 513, A74

\bibitem[\protect\citeauthoryear{Negueruela et al.}{2011}]{neg11} Negueruela, I., Gonz\'alez-Fern\'andez, C., Marco, A., \& Clark, J.S. 2011, \aap, 528, A59

\bibitem[\protect\citeauthoryear{Poelarends et al.}{2008}]{poel08} Poelarends, A.J.T., Herwig, F., Langer, N., \& Heger, A. 2008, ApJ, 675, 614

\bibitem[\protect\citeauthoryear{Salasnich, Bressan \& Chiosi}{Salasnich et al.}{1999}]{salas99} Salasnich, B., Bressan, A., \& Chiosi, C. 1999, A\&A, 342, 131

\bibitem[\protect\citeauthoryear{Shortridge et al.}{2004}]{shortridge} Shortridge, K., et al.\ 2004, Starlink User Note 86.21, R.A.L.

\bibitem[\protect\citeauthoryear{Sowell}{1984}]{sowella} Sowell, J.R. 1984, ApJS, 55, 455

\bibitem[\protect\citeauthoryear{Sowell}{1987}]{sowellb} Sowell, J.R. 1987, ApJS, 64, 241

\bibitem[Schmidt(1982)]{schmidt82} Schmidt, E.G. 1982, AJ, 87, 1197

\bibitem[\protect\citeauthoryear{Schmidt-Kaler}{1982}]{skaler82} Schmidt-Kaler, T., 1982, Landolt-B\"ornstein, N.S. VI, 2b

\bibitem[\protect\citeauthoryear{Skrutskie et al.} {2006}]{skru06} Skrutskie, M.F., 
et al. 2006, AJ, 131, 1163

\bibitem[\protect\citeauthoryear{Smartt}{2009}]{smartt09} Smartt, S.J. 2009, ARA\&A, 47, 63

\bibitem[\protect\citeauthoryear{Smartt et al.}{2009}]{smarttal} Smartt, S.J., Eldridge, J.J., Crockett, R.M., \& Maund, J.R. 2009, MNRAS, 395, 1409

\bibitem[\protect\citeauthoryear{Smith et al.}{2011}]{smith11} Smith, N., Li, W., Filippenko, A.V., \& Chornock, R. 2011, MNRAS, in press

\bibitem[\protect\citeauthoryear{Stetson}{1987}]{stetson1987} Stetson, P.B. 1987, PASP, 99, 191

\bibitem[Strai\v{z}ys \& Lazauskait\.{e}(2009)]{straizys09}
Strai\v{z}ys, V., \& Lazauskait\.{e}, R. 2009, Balt. Astr., 18, 19

\bibitem[\protect\citeauthoryear{Tadross}{2008}]{tadross} Tadross, A.L. 2008, MNRAS, 389, 285

\bibitem[\protect\citeauthoryear{Turner}{1980}]{turner80} Turner, D.G. 1980, PASP, 92, 840

\bibitem[\protect\citeauthoryear{Turner}{1981}]{turner81} Turner, D.G. 1981, AJ, 86, 222

\bibitem[\protect\citeauthoryear{Turner}{1992}]{turner92} Turner, D.G. 1992, AJ, 104, 1865

\bibitem[\protect\citeauthoryear{Turner}{1996}]{turner96} Turner, D.G. 1996, JRASC, 90, 82


\bibitem[\protect\citeauthoryear{Turner et al.}{1994}]{turner94} Turner, D.G., Mandushev, G.I., \& Forbes, D.  1994, AJ, 107, 1796

\bibitem[\protect\citeauthoryear{Turner et al.}{2008}]{turner08} Turner, D.G., et al. 2008, MNRAS, 388, 444

\bibitem[\protect\citeauthoryear{Valdes et al.}{2004}]{indous} Valdes, F., Gupta, R., Rose, J.A., Singh, H.P., \& Bell, D.J.
2004, ApJS, 152, 251 

\bibitem[\protect\citeauthoryear{Wegner}{1994}]{wegner} Wegner, W. 1994, MNRAS, 270, 229

\bibitem[Weidemann(2000)]{weide00} Weidemann, V. 2000, \aap, 363, 647

\bibitem[\protect\citeauthoryear{Winkler}{1997}]{winkler} Winkler, H. 1997, MNRAS, 287, 481

\bibitem[Yadav \& Sagar(2004)]{ys04} Yadav, R.K.S., \& Sagar, R. 2004, \mnras, 349, 1481

\bibitem[\protect\citeauthoryear{Zhou}{1991}]{zhou91} Zhou, X. 1991, A\&A, 248, 367

\end{thebibliography}
\end{document}